\documentclass[a4paper,12pt,reqno]{article}
\usepackage{amsmath}

\allowdisplaybreaks

\renewcommand{\i}{\mathrm{i}}
\newcommand{\al}{\alpha}
\newcommand{\ad}{\dot{\alpha}}
\newcommand{\be}{\beta}
\newcommand{\bd}{\dot{\beta}}
\newcommand{\de}{\delta}
\newcommand{\ep}{\varepsilon}
\newcommand{\si}{\sigma}
\newcommand{\bsi}{\bar{\sigma}}
\newcommand{\la}{\lambda}
\newcommand{\bla}{\bar{\lambda}}
\newcommand{\bpsi}{\bar{\psi}}
\newcommand{\half}{\tfrac{1}{2}}
\newcommand{\ihalf}{\tfrac{\mathrm{i}}{2}}
\newcommand{\quart}{\tfrac{1}{4}}
\newcommand{\tab}{\quad\,}
\newcommand{\bZ}{\bar{Z}}
\newcommand{\p}{\partial}
\newcommand{\D}{\mathcal{D}}
\newcommand{\Db}{\bar{\mathcal{D}}}
\newcommand{\com}[2]{[\,#1\, ,\,#2\,]}
\newcommand{\aco}[2]{\{#1\, ,\,#2\}}
\DeclareSymbolFont{AMSb}{U}{msb}{m}{n}
\DeclareMathSymbol{\fieldR}{\mathalpha}{AMSb}{"52}

\begin{document}

\begin{flushright}
 ITP-UH-13/98 \\ hep-th/9805199
\end{flushright}
\medskip

\begin{center}
 {\Large\bfseries The Linear Vector-Tensor Multiplet \\
	with Gauged Central Charge} \\[5mm]
 Norbert Dragon and Ulrich Theis \\[2mm]
 \textit{Institut f\"ur Theoretische Physik, Universit\"at Hannover \\
	Appelstra\ss{}e 2, 30167 Hannover, Germany}
\end{center}
\vspace{6mm}

\centerline{\bfseries Abstract} \medskip

We derive consistent superfield constraints for the linear vector-tensor
multiplet with gauged central charge. The central charge transformations
and the action turn out to be nonpolynomial in the gauge field.
\vspace{10mm}


The $N=2$ vector-tensor (VT) multiplet \cite{SSW} is a combination of an
$N=1$ vector- and an $N=1$ linear multiplet, and as such contains among its
components a 1-form and a 2-form gauge field. Since its rediscovery by string
theorists \cite{Louis,Siebelink}, much effort has been made to derive
consistent interactions both with itself \cite{Claus,Kuzenko,Ivanov} and with
background $N=2$ vector multiplets \cite{Claus,Grimm,Ovrut,DKT}. In
\cite{Claus} the VT multiplet was coupled to an abelian vector multiplet that
gauges the central charge of the $N=2$ supersymmetry algebra, resulting in an
interaction of three gauge fields.

Such interactions of higher degree gauge fields have recently been classified
by cohomological means (outside the framework of supersymmetry) in
\cite{Henneaux}. As a particular example an interaction of two 1-forms and a
2-form was studied in \cite{Brandt} which turned out to be nonpolynomial and
resembled the bosonic sector of the VT multiplet with gauged central charge.

In the present letter, we show that this model is indeed contained in the
linear VT multiplet with gauged central charge, which is derived from a set
of constraints on the basic superfield, thus providing the ground for a
superspace formulation of the component results of \cite{Claus}. Here we
follow an approach already applied in \cite{Kuzenko,Ivanov}, where the
nonlinear VT multiplet (with global central charge) was formulated in
harmonic superspace, namely to determine consistent deformations of the flat
superfield constraints.
\medskip

Our considerations are based on the $N=2$ supersymmetry algebra
 \begin{align} \label{algebra}
  \aco{\D_\al^i}{\Db_{\ad j}} & = -\i \de_j^i\, \si^a_{\al\ad} \D_a &
	\com{\D_a}{\D_b} & = F_{ab}\, \de_z \notag \\
  \aco{\D_\al^i}{\D_\be^j} & = \ep_{\al\be}\, \ep^{ij} \bZ \de_z &
	\com{\D_\al^i}{\D_a} & = \ihalf (\si_a \bla^i)_\al \de_z \\
  \aco{\Db_{\ad i}}{\Db_{\bd j}} & = \ep_{\ad\bd}\, \ep_{ij} Z \de_z &
	\com{\Db_{\ad i}}{\D_a} & = \ihalf (\la_i \si_a)_{\ad} \de_z\ .
	\notag
 \end{align}
The real generator $\de_z$ is central and commutes with all generators of
the supersymmetry algebra,
 \begin{equation}
  \de_z = \de_z{}^\ast\, ,\quad \com{\de_z}{\D_\al^i} = 0\, ,\quad
	\com{\de_z}{\Db_{\ad j}} = 0\, ,\quad \com{\de_z}{\D_a} = 0\, .
 \end{equation}
$\D_a$ is the gauge covariant derivative
 \begin{equation} \label{deriv}
  \D_a = \p_a + A_a \de_z
 \end{equation}
and $F_{ab}$ the abelian field strength of the gauged central charge
 \begin{equation}
  F_{ab} = \p_a A_b - \p_b A_a\, .
 \end{equation}
The algebra \eqref{algebra} might as well be read as an algebra with a
gauged abelian transformation and no central charge. However, the scalar
field $Z$ has to have a nonvanishing constant background value as we will
see.

$Z$, $\bZ$, $\la_{\underline{\al}}^i$ and $A_a$ are component fields of the
vector multiplet which is completed by an auxiliary triplet $Y^{ij}=Y^{ji}
=(Y_{ij})^\ast$,
 \begin{equation} \begin{split}
  \la_\al^i & = \D_\al^i Z\, ,\quad \bla_{\ad i} = \Db_{\ad i} \bZ\, , \\
  Y^{ij} & = \half \D^i \D^j Z = \half \Db^i \Db^j \bZ\, , \\
  F_{ab} & = \quart (\D^i \si_{ab} \D_i Z - \Db_i \bsi_{ab} \Db^i \bZ)\, .
 \end{split} \end{equation}
They are tensor fields and carry no central charge. So they are unchanged by
gauged central charge transformations $s_z$ with a real gauge parameter
$C(x)$ with arbitrary spacetime dependence. Under these transformations the
vector field $A_a$ is changed by the gradient of the gauge parameter,
 \begin{equation} \label{trans}
  s_z Z = 0\, ,\quad s_z \la_{\underline{\al}}^i = 0\, ,\quad s_z Y^{ij}
	= 0\, ,\quad s_z A_a = -\p_a C\, .
 \end{equation}

To construct invariant actions, we make use of the linear multiplet with
gauged central charge as in \cite{vanHolten}. One starts from a field
 \begin{equation}
  \mathcal{L}^{ij} = \mathcal{L}^{ji} = (\mathcal{L}_{ij})^\ast
 \end{equation}
satisfying the constraints
 \begin{equation} \label{linear}
  \D_\al^{(i} \mathcal{L}^{jk)} = 0 = \Db_{\ad}^{(i} \mathcal{L}^{jk)}\, .
 \end{equation}
It contains among its components a real vector $V^a = - \frac{\i}{6} \D_i
\si^a \Db_j \mathcal{L}^{ij}$, which is con\-strained by
 \begin{equation}
  \D_a V^a = \frac{1}{12} \de_z  \big[ Z \D_i \D_j + 3 Y_{ij} + 4 \la_i
	\D_j \big] \mathcal{L}^{ij} + \text{h.c.}
 \end{equation}
{}From the identity
 \begin{equation*}
  s_z (A_a V^a) = - (\p_a C) V^a + A_a C \de_z V^a = - \p_a (C V^a) +
	C \D_a V^a
 \end{equation*}
it follows immediately that
 \begin{equation} \label{L1}
  \mathcal{L} = \frac{1}{12} \big[ Z \D_i \D_j + 3 Y_{ij} + 4 \la_i \D_j
	+ \i A_a \D_i \si^a \Db_j \big] \mathcal{L}^{ij} + \text{h.c.}
 \end{equation}
is invariant under gauged central charge transformations upon integration
over spacetime. This action is also $N=2$ supersymmetric, as can be checked
by explicit calculation. So once we have constructed a multiplet on which
the algebra \eqref{algebra} is realized we try to construct a composite
field $\mathcal{L}^{ij}$ which satisfies \eqref{linear} and use \eqref{L1}
as the Lagrangian.
\medskip


For the VT multiplet gauging the central charge is quite involved.
This multiplet is obtained from a real scalar field $L$, which for ungauged
central charge satisfies the constraints
 \begin{equation} \label{flat}
  L = L^\ast\, ,\quad D_\al^{(i} \bar{D}_{\ad}^{j)} L = 0\, ,\quad
	D^{(i} D^{j)} L = 0\, .
 \end{equation}
Simply replacing the flat derivatives $D_\al^i$ and $\bar{D}_{\ad}^j$ with
gauge covariant derivatives $\D_\al^i$ and $\Db_{\ad}^j$ leads to
inconsistencies that show up at a rather late stage in the process of
evaluating the algebra on the multiplet components (see \cite{DIKST} for
details). This problem can be avoided, however, by a suitable change of the
above constraints. They must preserve the field content as compared to the
case of ungauged central charge. A consistent set of constraints for the
VT multiplet with gauged central charge is given by
 \begin{equation} \begin{split} \label{constraints}
  \D_\al^{(i} \Db_{\ad}^{j)} L & = 0\, , \\
  \D^{(i} \D^{j)} L & = \frac{2}{\raisebox{-1pt}{$\bZ - Z$}} \big( \D^{(i}
	Z \D^{j)} L + \Db^{(i} \bZ \Db^{j)} L + \half L \D^{(i} \D^{j)} Z
	\big)\, .
 \end{split} \end{equation}
Obviously, this requires the imaginary part of $Z$ to have a nonvanishing
background value. For $Z = \i$ the constraints reduce to the flat case, eq.\
\eqref{flat}.

Let us show how these constraints were obtained. We considered the following
Ansatz, with the coefficients being arbitrary functions of $Z$ and $\bZ$,
 \begin{equation} \begin{split} \label{ansatz}
  \D_\al^{(i} \Db_{\ad}^{j)} L & = a\, \D_\al^{(i} Z \Db_{\ad}^{j)} L
	+ \bar{a}\, \D_\al^{(i} L \Db_{\ad}^{j)} \bZ + b L\, \D_\al^{(i}
	Z \Db_{\ad}^{j)} \bZ\, ,\quad \text{$b$ real}, \\
  \D^{(i} \D^{j)} L & = A\, \D^{(i} Z \D^{j)} L + B\, \Db^{(i} \bZ \Db^{j)}
	L + C L\, \D^{(i} \D^{j)} Z \\*
  & \tab + D L\, \D^{(i} Z \D^{j)} Z + E L\, \Db^{(i} \bZ \Db^{j)} \bZ\, .
 \end{split} \end{equation}
It is linear in the components of the VT multiplet and hence
invariant under a rescaling of $L$ with a constant parameter. The
constraints have to satisfy the necessary consistency conditions
 \begin{equation} \label{con0}
  \D_\al^{(i} \D^j \D^{k)} L = 0\, ,\quad \Db_{\ad}^{(i} \D^j \D^{k)} L
	= \D^{(i} \D^j \Db_{\ad}^{k)} L\, ,
 \end{equation}
which yield a system of differential equations for the coefficient
functions. The first condition requires
 \begin{equation} \begin{split} \label{con1}
  0 & = C - \half A \\
  0 & = \p_Z C - D - \half A C \\
  0 & = E + \bar{a} B \\
  0 & = \p_Z E - \half A E + b B \\
  0 & = \p_Z B + B (a - \half A) \\
  0 & = \p_Z A - \half A^2 - 2 D\, ,
 \end{split} \end{equation}
while the second leads to
 \begin{equation} \begin{split} \label{con2}
  0 & = C - \half B - a \\
  0 & = \p_{\bZ} C - E - \half B \bar{C} - \bar{a} C - b \\
  0 & = \p_Z a + a (a - A) - D \\
  0 & = \p_{\bZ} D - \p_Z b - \half B \bar{E} - \bar{a} D - b (a - A) \\
  0 & = \p_{\bZ} A - \p_Z \bar{a} - \half B \bar{B} - a \bar{a} - b \\
  0 & = \p_{\bZ} B - B (\bar{a} + \half \bar{A}) - 2 E\, .
 \end{split} \end{equation}
If we perform a more general rescaling $L = f(Z, \bZ)\, \hat{L}$ in
\eqref{ansatz}, the coefficients transform as
 \begin{gather}
  \hat{a} = a - \p_Z f / f\, ,\quad \hat{b} = (b f + a \p_{\bZ} f +
	\bar{a} \p_Z f - \p_Z \p_{\bZ} f) / f\, , \notag \\
  \hat{A} = A - 2 \p_Z f / f\, ,\quad \hat{B} = B\, ,\quad \hat{C} =
	C - \p_Z f / f\, , \\
  \hat{D} = (D f + A \p_Z f - \p_Z^2 f) / f\, ,\quad \hat{E} = (E f
	+ B \p_{\bZ} f) / f\, . \notag
 \end{gather}
We can choose $f$ such that $\hat{a}$ vanishes, which simplifies the
conditions \eqref{con1} and \eqref{con2} considerably, and obtain (omitting
the hats)
 \begin{equation}
  2 C = B = A\, ,\quad a = b = D = E = 0\, ,
 \end{equation}
where $A$ has to satisfy the differential equations
 \begin{equation}
  \p_Z A = \half A^2\, ,\quad \p_{\bZ} A = \half A \bar{A}\, .
 \end{equation}
Their general (nonvanishing) solution is
 \begin{equation}
  A(Z,\bZ) = \frac{2\, \mathrm{e}^{-\i\varphi}}{\raisebox{-1pt}{$
	\mathrm{e}^{\i\varphi} \bZ - \mathrm{e}^{-\i\varphi} Z
	+ 2\i\, r$}}\, ,\quad r, \varphi \in \fieldR\, .
 \end{equation}
The parameters can be removed by a redefinition
 \begin{equation}
  Z \rightarrow \mathrm{e}^{\i\varphi} (Z + \i r)\, ,\quad \D_\al^i
	\rightarrow \mathrm{e}^{-\i\varphi/2} \D_\al^i\, ,
 \end{equation}
which eventually leads to the constraints given in eq.\ \eqref{constraints}.
Of course, the conditions \eqref{con0} do not guarantee the consistency of
these constraints, we now have to carefully evaluate the algebra on each
component field.

The linearity of the constraints \eqref{constraints} in $L$ and its
supersymmetry transformations implies that we will not encounter any
self-interactions if $\mathcal{L}^{ij}$ is quadratic in $L$. To find
constraints that underlie the nonlinear VT multiplet as described in
\cite{Claus}, one has to start from a more general Ansatz.
\medskip


We now investigate the consequences of the constraints \eqref{constraints}.
The independent components of the multiplet can be chosen as
 \begin{gather}
  L\, ,\quad \psi_\al^i = \i \D_\al^i L\, ,\quad \bpsi_{\ad i} = -\i
	\Db_{\ad i} L\, ,\quad  U = \de_z L\, , \notag \\
  G_{\al\be} = \half \com{\D_\al^i}{\D_{\be i}} L\, ,\quad \bar{G}_{\ad\bd}
	= \half \com{\Db_{i\ad}}{\Db_{\bd}^i} L\, , \\
  W_{\al\ad} = -\half \com{\D_\al^i}{\Db_{\ad i}} L\, . \notag
 \end{gather}
In addition some abbreviations will prove useful in the following,
 \begin{equation} \begin{split}
  I = \mathrm{Im}\, Z\, , & \quad R = \mathrm{Re}\, Z\, , \\
  \Lambda_a = L \p_a R + \half (\la^i \si_a \bpsi_i + \psi^i \si_a \bla_i
	)\, , & \quad \Sigma_{ab} = L F_{ab} + \i (\la_i \si_{ab} \psi^i
	- \bpsi_i \bsi_{ab} \bla^i)\, .
 \end{split} \end{equation}
$R$, $\Lambda_a$ and $\Sigma_{ab}$ vanish in the case of a global central
charge, i.e.\ when $Z$ is reduced to its background value $Z = \i$.

{}From the algebra \eqref{algebra} we obtain the supersymmetry
transformations
 \begin{equation} \begin{split}
  \D_\al^i \psi_\be^j & = \ihalf \ep^{ij} (\ep_{\al\be} \bZ U + \si^{ab}
	{}_{\al\be} G_{ab}) + \frac{\i}{2I} \ep_{\al\be} (\la^{(i}
	\psi^{j)} - \bla^{(i} \bpsi^{j)} + \i Y^{ij} L)\, , \\
  \Db_{\ad}^i \psi_\al^j & = \half \ep^{ij} \si^a_{\al\ad} (\D_a L + \i
	W_a)\, , \\
  \D_\al^i W_a & = \big( \i \bZ\, \si_a \de_z \bpsi^i + \half U \si_a
	\bla^i - \D_a \psi^i \big)_\al\, , \\
  \D_\al^i G_{ab} & = \big( 2I\, \si_{ab} \de_z \psi^i + U \si_{ab} \la^i
	+ \i \ep_{abcd} \si^c \D^d \bpsi^i \big)_\al\, ,
 \end{split} \end{equation}
and the central charge transformations. These contain covariant derivatives
which in turn again contain $\de_z$, eq.\ \eqref{deriv}, so the
transformations are given only implicitly. One could solve for $\de_z$ of the
fields, however at this stage it is more advantageous to use the manifestly
covariant expressions (here a tilde denotes the dual of a 2-form, i.e.\
$\tilde{G}^{ab} = \half \ep^{abcd} G_{cd}$)
 \begin{align}
  \begin{split}
  \de_z \psi^i & = \frac{\i}{Z} \big( \si^a \D_a \bpsi^i - \la^i U \big)
	+ \frac{\i}{2ZI} \Big[ \ihalf \bZ \la^i U - Y^{ij} \psi_j + \i
	\si^a \bpsi^i \p_a \bZ \\*
  & \tab + L \si^a \p_a \bla^i + \half (\D_a L - \i W_a) \si^a \bla^i -
	F_{ab} \si^{ab} \psi^i - \ihalf G_{ab} \si^{ab} \la^i \\*
  & \tab - \frac{\i}{2I} \la_j (\la^{(i} \psi^{j)} - \bla^{(i} \bpsi^{j)}
	+ \i Y^{ij} L) \Big]\, ,
  \end{split} \notag \\
  \de_z \big( I & W_a - \Lambda_a \big) = \D^b G_{ab}\, , \\
  \de_z \big( I & \tilde{G}_{ab} + R G_{ab} + \Sigma_{ab} \big) = -
	\ep_{abcd} \D^c W^d\, . \notag
 \end{align}

As in the case of the ordinary VT multiplet, the vector $W_a$ and the
antisymmetric tensor $G_{ab}$ are subject to some constraints. Their
solvability is the final consistency check. Closure of the algebra requires
 \begin{equation} \begin{split}
  \D^a & \big( I W_a - \Lambda_a \big) = \half F_{ab} G^{ab}\, , \\
  \D^a & \big( I \tilde{G}_{ab} + R G_{ab} + \Sigma_{ab} \big) =
	\tilde{F}_{bc} W^c\, .
 \end{split} \end{equation}
With the help of the central charge transformations we obtain from these
covariant expressions the equations
 \begin{equation} \begin{split} \label{BI}
  \p^a & \big( I W_a - G_{ab} A^b - \Lambda_a \big) = 0\, , \\
  \p^a & \big( I \tilde{G}_{ab} + R G_{ab} + \ep_{abcd} A^c W^d +
	\Sigma_{ab} \big) = 0\, ,
 \end{split} \end{equation}
which identify the brackets as duals of the field strength of a 2-form
$B_{ab}$ and a 1-form $V_a$, respectively,
 \begin{gather}
  I W^a = \half \ep^{abcd} (\p_b B_{cd} - A_b \tilde{G}_{cd}) + \Lambda^a\,
	, \label{IW} \\
  I \tilde{G}_{ab} + R G_{ab} = \ep_{abcd} (\p^c V^d - A^c W^d) - \Sigma_{ab}
	\, . \label{IG}
 \end{gather}

We cannot read off $W_a$ and $G_{ab}$, however, since the equations are
coupled. Solving eq.\ \eqref{IG} for $G_{ab}$ and inserting this into eq.\
\eqref{IW}, we find
 \begin{equation}
  I (E \delta^a_b + A^a A_b) W^b = |Z|^2 (H^a + T^{ab}\! A_b)\, ,
 \end{equation}
where we used the abbreviations
 \begin{equation} \begin{split}
  E & = |Z|^2 - A^a A_a\, , \\
  H^a & = \half \ep^{abcd}\, \p_b B_{cd} + \Lambda^a\, , \\
  T_{ab} & = \frac{I}{|Z|^2} \big( \p_a V_b - \p_b V_a + \tilde{\Sigma}_{ab}
	\big) + \frac{R}{|Z|^2} \big( \ep_{abcd} \p^c V^d - \Sigma_{ab}
	\big)\, .
 \end{split} \end{equation}
The inverse of the matrix $(E \delta^a_b + A^a A_b)$ is
 \begin{equation}
  \frac{1}{E} (\delta^b_a - |Z|^{-2} A^b A_a)\, ,
 \end{equation}
which gives
 \begin{equation}
  W^a = \frac{1}{IE} \big( |Z|^2 H^a -  A^a A_b H^b + |Z|^2 T^{ab}\!
	A_b \big)\, ,
 \end{equation}
and finally, from eq.\ \eqref{IG},
 \begin{equation}
  G_{ab} = T_{ab} - \frac{2}{E} A_{[a} \big( H_{b]} + T_{b]c} A^c
	\big) - \frac{R}{IE} \ep_{abcd} A^c \big( H^d + T^{de}\! A_e
	\big)\, .
 \end{equation}
These expressions are nonpolynomial in the gauge field $A_a$ due to the
appearance of $E$ in the denominator. Although it appears as if they would
transform inhomogeneously under the central charge transformation $s_z$,
explicit calculation shows that the gradient of the gauge parameter, eq.\
\eqref{trans}, cancels, so $W_a$ and $G_{ab}$ are indeed tensors as assumed
from the beginning.

It remains to give the transformations of the potentials $V_a$ and $B_{ab}$.
As the brackets in \eqref{IW} and \eqref{IG} already indicate, the central
charge transformations read
 \begin{equation}
  \de_z V_a = - W_a\, ,\quad \de_z B_{ab} = - \tilde{G}_{ab}\, ,
 \end{equation}
while the supersymmetry generators act as
 \begin{equation} \begin{split}
  \D_\al^i V_a & = \big( A_a \psi^i - \half L \si_a \bla^i - \i \bZ \si_a
	\bpsi^i \big)_\al\, , \\
  \D_\al^i B_{ab} & = - 2\i\, \big( I \si_{ab} \psi^i + \half L \si_{ab}
	\la^i + A_{[a} \si_{b]} \bpsi^i \big)_\al\, .
 \end{split} \end{equation}
Then the algebra closes up to gauge transformations of the form
 \begin{equation}
  V_a \rightarrow V_a + \p_a \xi\, ,\quad B_{ab} \rightarrow B_{ab} +
	\p_a \xi_b - \p_b \xi_a
 \end{equation}
when evaluated on the potentials.
\medskip


Next, we construct a linear multiplet $\mathcal{L}^{ij}$. Given the
constraints \eqref{constraints} on $L$ it is easy to find a field with the
required properties (note that $\D^{(i} \D^{j)} L$ is imaginary),
 \begin{equation} \begin{split}
  \mathcal{L}^{ij} & = \i \big( \Db^{(i} L \Db^{j)} L - \D^{(i} L \D^{j)} L
	- L \D^{(i} \D^{j)} L \big) \\
  & = \i (\psi^i \psi^j - \bpsi^i \bpsi^j) - \frac{\i}{I} L \big(
	\la^{(i} \psi^{j)} - \bla^{(i} \bpsi^{j)} + \i Y^{ij} L \big)\, .
 \end{split} \end{equation}
Applying the rule \eqref{L1} we eventually obtain the Lagrangian
 \begin{equation} \begin{split}
  \mathcal{L} & = \half I \big( \p^a\! L\, \p_a L - W^a W_a + (|Z|^2 - A^a
	A_a) U^2 \big) - \half L^2 \p^a \p_a I \\*
  & \tab - \quart G^{ab} \big( I G_{ab} - R \tilde{G}_{ab} + 4 A_{[a} W_{b]}
	\big) - \frac{1}{4I} Y^{ij} Y_{ij} L^2 \\*
  & \tab + \text{fermion terms}\, ,
 \end{split} \end{equation}
where we used the identities \eqref{BI} to combine terms into a total
derivative which was dropped thereafter. To compare with the action found
in \cite{Brandt}, we insert the expressions for $W_a$ and $G_{ab}$,
 \begin{equation} \begin{split}
  \mathcal{L} & = \half I \p^a\! L\, \p_a L - \half L^2 \p^a \p_a I + \half
	I E\, U^2 - \frac{1}{4I} Y^{ij} Y_{ij} L^2 \\*
  & \tab - \quart T^{ab} (I T_{ab} - R \tilde{T}_{ab}) - \frac{|Z|^2}{2IE}
	(H^a + T^{ab}\! A_b)^2 + \frac{1}{2IE} (A_a H^a)^2 \\*
  & \tab + \text{fermion terms}\, .
 \end{split} \end{equation}

With $L$ and $U$ set to zero and $Z = \i$, which breaks supersymmetry but
keeps central charge invariance, the bosonic Lagrangian indeed coincides
with the one given in ref.\ \cite{Brandt}:
 \begin{equation}
  \mathcal{L} = - \frac{1}{4} T^{ab} T_{ab} - \frac{1}{2}\, \frac{(H^a +
	T^{ac}\! A_c)^2}{1 - A^b A_b} + \frac{1}{2}\, \frac{(A_a H^a)^2}{1
	- A^b A_b} - \frac{1}{4g^2} F^{ab} F_{ab}\, ,
 \end{equation}
where now $H^a = \half \ep^{abcd} \p_b B_{cd}$ and $T_{ab} = \p_a V_b - \p_b
V_a$, and a kinetic term for $A_a$ has been added. The second term contains
a vertex involving all three gauge fields $A_a$, $V_a$ and $B_{ab}$, which
has been identified as a Freedman-Townsend coupling in \cite{Henneaux}.
\bigskip


\textbf{Acknowledgments} \\
We thank F.~Brandt, P.~Fayet and S.~Kuzenko for stimulating and enjoyable
discussions.

\end{document}